\documentclass{elsart}
\usepackage{amssymb}
\usepackage{graphicx}
\usepackage{amsmath}
\usepackage{cite}
\usepackage{bm}
\usepackage{float}

\begin{document}
\begin{frontmatter}
\title{Metal-insulator transition and low-density phases in a strongly-interacting two-dimensional electron system}
\author{A.~A. Shashkin}
\address{Institute of Solid State Physics, Chernogolovka, Moscow District 142432, Russia}
\author{S.~V. Kravchenko\corauthref{cor1}}
\address{Physics Department, Northeastern University, Boston, Massachusetts 02115, USA}
\corauth[cor1]{Corresponding author}
\ead{s.kravchenko@northeastern.edu}

\begin{abstract}
We review recent experimental results on the metal-insulator transition and low-density phases in strongly-interacting, low-disordered silicon-based two-dimensional electron systems. Special attention is given to the metallic state in ultra-clean SiGe quantum wells and to the evidence for a flat band at the Fermi level and a quantum electron solid.
\end{abstract}

\end{frontmatter}

\section{Introduction}\vspace{-5mm}
\enlargethispage{5mm}
The metal-insulator transition (MIT) is an exceptional testing ground for studying strong electron-electron correlations in two dimensions (2D) in the presence of disorder. The existence of the metallic state and the MIT in strongly interacting 2D electron systems (contrary to the famous conclusion by the ``Gang of Four'' that only an insulating state is possible in non-interacting 2D systems \cite{abrahams1979scaling}) was predicted in Refs.~\cite{finkelstein1983influence,finkelstein1984weak,castellani1984interaction}. The phenomenon was experimentally discovered in silicon metal-oxide-semiconductor field-effect transistors (MOSFETs) and subsequently observed in a wide variety of other strongly-interacting 2D systems: $p$- and $n$-SiGe heterostructures, $p$- and $n$-GaAs/AlGaAs hetero\-struc\-tures, AlAs heterostructures, ZnO-related hetero\-struc\-tures, \textit{etc}.\ (for reviews, see Refs.~\cite{sarachik1999novel,abrahams2001metallic,kravchenko2004metal,shashkin2005metal,pudalov2006metal,spivak2010transport,kravchenko2017strongly,shashkin2017metal,shashkin2019recent,dolgopolov2019two} and references therein). Now it is widely accepted that the driving force behind the MIT is the strong correlations between carriers. Here we review recent progress in the studies of the MIT and related phenomena.  Section 2 is devoted to the MIT in zero magnetic field and behavior of the effective electron mass in ultra-high mobility SiGe/Si/SiGe quantum wells.  In section 3, we show that in an exceptionally clean two-valley system, the metallic state survives even when the spins of the electrons become completely polarized.  Section 4 describes the scaling analysis of the temperature dependences of the resistance in the spirit of the dynamical mean-field theory and renormalization-group theory.  In section 5, the formation of the flat band at the Fermi level in SiGe/Si/SiGe quantum well is discussed.  Finally, transport evidence for the formation of the quantum electron solid in silicon MOSFETs at very low electron densities is presented in section 6.

We used two sets of samples.  The first set consisted of CVD-grown ultraclean SiGe/Si/SiGe quantum wells described in detail in Refs.~\cite{melnikov2015ultra,melnikov2017unusual}. The maximum electron mobility, $\mu$, in these samples reaches 240~m$^2$/Vs. The approximately 15~nm wide silicon (001) quantum well was sandwiched between Si$_{0.8}$Ge$_{0.2}$ potential barriers. The samples were patterned in Hall-bar shapes using standard photo-lithography; the distance between the potential probes and the width were 150~$\mu$m and 50~$\mu$m, correspondingly. The second set was (100)-silicon MOSFETs with a peak electron mobility of 3~m$^2$/Vs similar to those described in Ref.~\cite{heemskerk1998nonlinear}. Samples had a Hall bar geometry of width 50~$\mu$m and distance between the potential probes of 120~$\mu$m. In both sets of samples, the electron density was controlled by applying a positive dc voltage to the gate relative to the contacts.

Measurements were carried out in Oxford TLM-400 and Kelvinox-100 dilution refrigerators. On the metallic side of the transition, the data were taken by a standard four-terminal lock-in technique in a frequency range 0.5--11~Hz in the linear regime of response. On the insulating side, the resistance was measured with {\it dc} technique using an electrometer with a high input impedance.

\section{Quantum phase transition in ultrahigh mobility SiGe/Si/SiGe two-dimensional electron system in zero magnetic field}\vspace{-5mm}

An important characteristic that defines the MIT is the magnitude of the resistance drop with decreasing temperature on the metallic side of the transition. Until recently, the strongest drop of the resistance (up to a factor of 7) was reported in clean silicon MOSFETs. At the same time, in much less disordered GaAs-based structures, the resistance drop has not exceeded a factor of about three. This discrepancy has been attributed primarily to the fact that electrons in silicon-based structures have two almost degenerate valleys in the energy spectrum, which strengthens the effects of correlations \cite{punnoose2001dilute,punnoose2005metal}.

\begin{figure}
\scalebox{.7}{\includegraphics{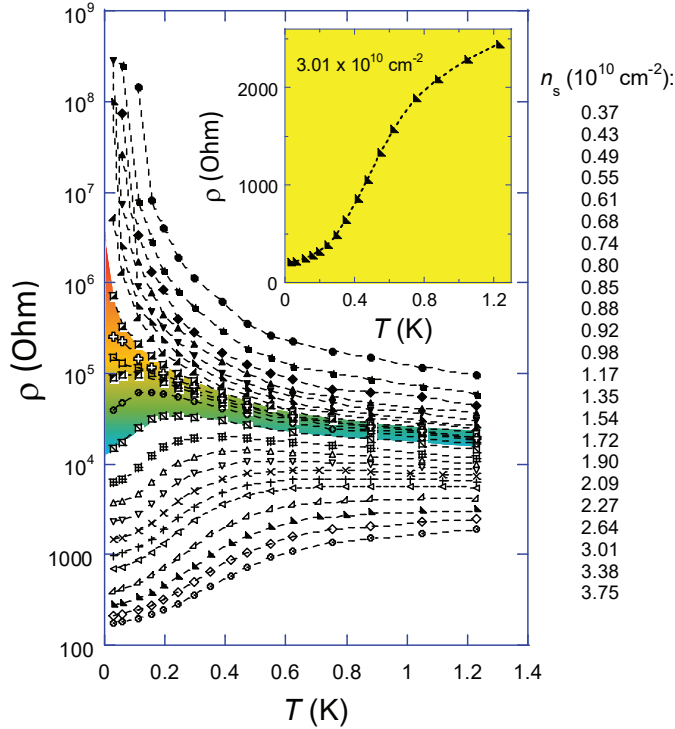}}
\caption{Temperature dependences of the resistivity in an ultralow-disorder SiGe/Si/SiGe quantum well at different electron densities in zero magnetic field.  Curves near the MIT are marked by the color-gradated area. In the inset, a close-up view of $\rho(T)$ displaying a drop of the resistivity by more than a factor of 12 is shown. From Ref.~\cite{melnikov2019quantum}.\label{fig1}}
\end{figure}

Resistivity $\rho$ of an ultraclean SiGe/Si/SiGe quantum well was measured in Ref.~\cite{melnikov2019quantum} as a function of temperature $T$ in a wide range of electron densities $n_{\text{s}}$, spanning both sides of the zero-magnetic-field MIT.  The data are plotted in Fig.~\ref{fig1}. At the highest temperature, the resistivity measured at the lowest electron density exceeds that at the highest density by less than two orders of magnitude, while at the lowest temperature, this difference becomes more than six orders of magnitude. The metal-insulator transition occurs at $n_{\text{c}}=0.88\pm0.02\times10^{10}$~cm$^{-2}$, according to the criterion of the sign change of the derivative ${\rm d}\rho/{\rm d}T$ (taking account of the tilted separatrix \cite{punnoose2005metal}).

The critical density determined in this way is almost an order of magnitude lower than that in the cleanest Si MOSFETs, where it is equal to $\approx 8\times10^{10}$~cm$^{-2}$. This difference can indeed be expected for an MIT driven by interactions. In Si MOSFETs, the value of the interaction parameter $r_{\text{s}}$, defined as the ratio between the Coulomb and Fermi energies, $r_{\text{s}}=g_{\text{v}}/(\pi n_{\text{s}})^{1/2}a_{\text{B}}$, reaches $\approx20$ at the critical electron density (here $g_{\text{v}}=2$ is the valley degeneracy and $a_{\text{B}}$ is the effective Bohr radius in semiconductor). SiGe/Si/SiGe quantum wells differ from Si MOSFETs by the strength of the disorder potential, the thickness of the 2D layer, and the dielectric constant equal to 7.7 in Si MOSFETs and to 12.6 in SiGe/Si/SiGe quantum wells. Since the dielectric constant is higher in the latter system, the interaction parameter is smaller by a factor of approximately 1.6 at the same electron density. The effective $r_{\text{s}}$ value is further reduced in the SiGe/Si/SiGe quantum wells due to the greater thickness of the 2D layer, which results in a smaller form-factor \cite{ando1982electronic}. We assume that the effective mass in the SiGe barrier is $\approx0.5\, m_{\text{e}}$ and estimate the barrier height at $\approx25$~meV. Evaluating the penetration of the wave function into the barrier, we obtain the effective thickness of the 2D layer to be $\approx200$~\AA\ compared to $\approx50$~\AA\ in Si MOSFETs. This results in the additional suppression of $r_{\text{s}}$ in SiGe/Si/SiGe quantum wells by a factor of about 1.3 with respect to Si MOSFETs. Thus, at the critical electron densities, the interaction parameters are close to 20 in both 2D systems, which is consistent with the results of Ref.~\cite{shashkin2007strongly}.

\begin{figure}
\scalebox{.5}{\includegraphics{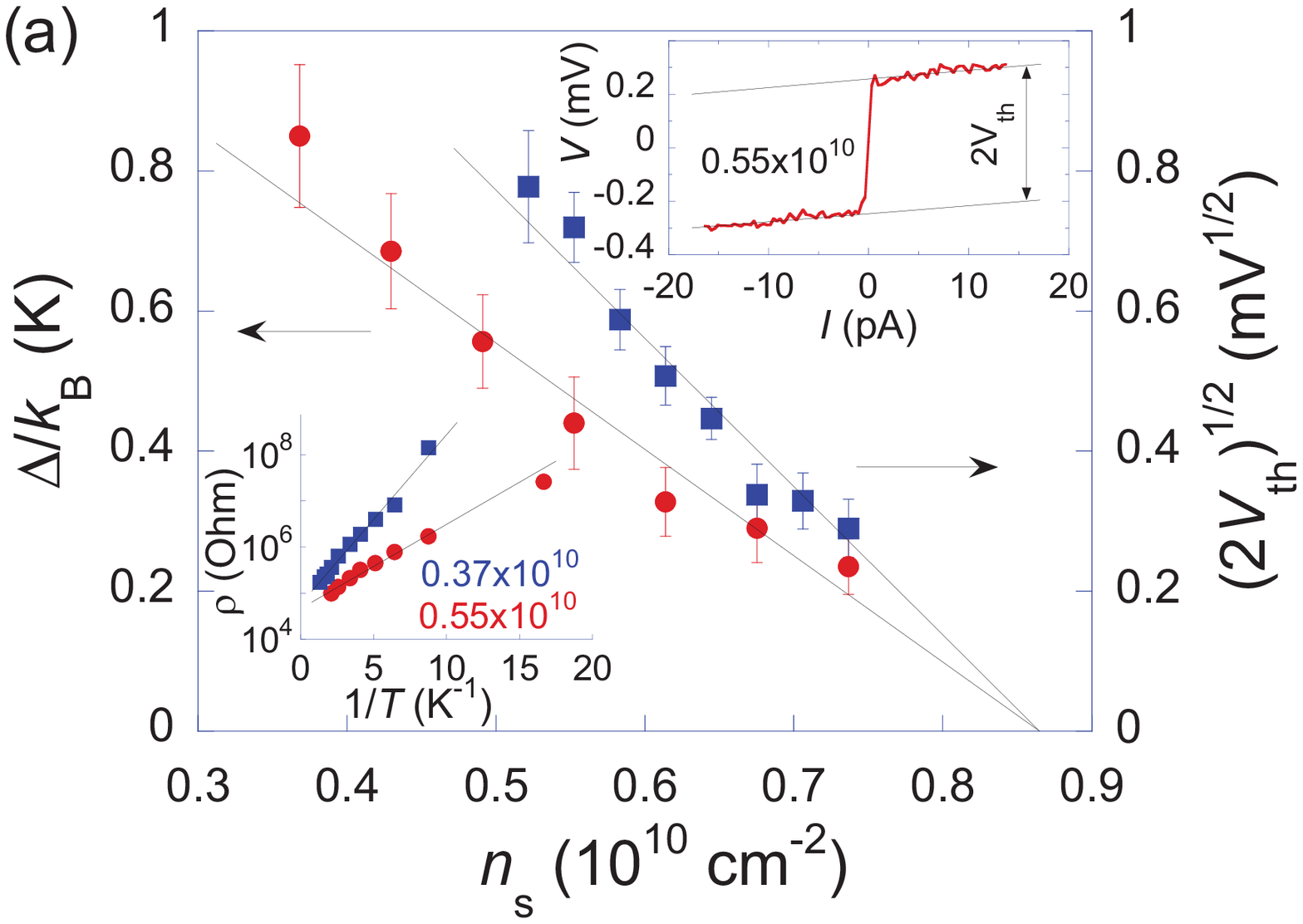}}\\
\scalebox{.5}{\includegraphics{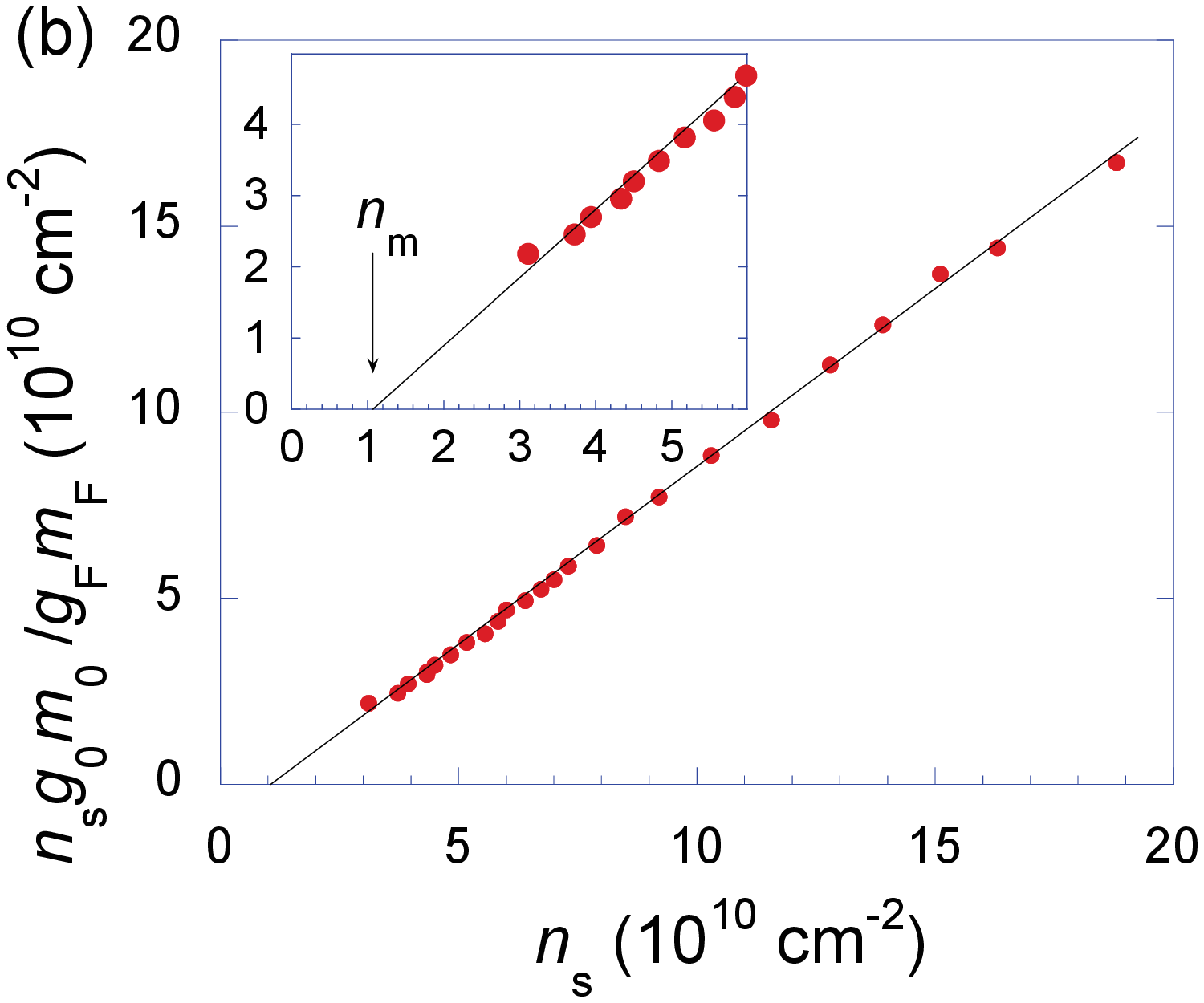}}
\caption{(a)~Activation energy and the square root of the threshold voltage as a function of the electron density in zero magnetic field. Vertical error bars correspond to the experimental uncertainty. The solid lines are linear fits yielding $n_{\text{c}}=0.87\pm0.02\times10^{10}$~cm$^{-2}$. Top inset: Current-voltage characteristic measured at a temperature of 30~mK in zero magnetic field. Bottom inset: Arrhenius plots of the resistivity in the insulating phase for two electron densities. The densities in both insets are indicated in cm$^{-2}$. (b)~Dependence of the effective mass at the Fermi level, $m_{\text{F}}$, on the electron density. The solid line is a linear fit. The experimental uncertainty corresponds to the data dispersion. The inset shows a close-up view of the dependence at low electron densities, where $n_{\text{m}}=1.1\pm0.1\times10^{10}$~cm$^{-2}$. From Ref.~\cite{melnikov2019quantum}.\label{fig2}}
\end{figure}

An alternative way to determine the critical density of the MIT is to study the insulating side of the transition.  The resistance has an activated form there, as shown in the bottom inset to Fig.~\ref{fig2}(a). In the main panel of Fig.~\ref{fig2}(a), the activation energy in temperature units, $\Delta/k_{\text{B}}$, is plotted \textit{vs}.\ the electron density (red circles). This dependence corresponds to the constant thermodynamic density of states near the critical point and is expected to be linear. The activation energy extrapolates to zero at $n_{\text{c}}=0.87\pm0.02\times10^{10}$~cm$^{-2}$, which matches, within the experimental uncertainty, the critical electron density determined from the temperature derivative criterion used above. Additionally, a typical low-temperature $I$-$V$ curve on the insulating side of the transition is a step-like function: the voltage abruptly rises at low currents and almost saturates at higher currents, as shown in the top inset to Fig.~\ref{fig2}(a). The magnitude of the step is $2\, V_{\text{th}}$, where $V_{\text{th}}$ is the threshold voltage. In Ref.~\cite{shashkin1994insulating}, such a threshold behavior of the $I$-$V$ characteristics has been attributed to the breakdown of the insulating phase that occurs when the localized electrons at the Fermi level acquire enough energy to reach the mobility edge in an electric field $V_{\text{th}}/d$ over a distance of the temperature-independent localization length $L$ (here $d$ is the distance between the potential probes). The values $\Delta/k_{\text{B}}$ and $V_{\text{th}}$ are related through the localization length, which diverges near the transition as $L(E_{\text{F}})\propto (E_{\text{c}}-E_{\text{F}})^{-s}$ with the exponent $s$ close to unity \cite{shashkin1994insulating} (here $E_{\text{c}}$ is the mobility edge and $E_{\text{F}}$ is the Fermi level). Therefore, the square root of $V_{\text{th}}$ should be a linear function of $n_{\text{s}}$ near the MIT, as indeed seen in Fig.~\ref{fig2}(a) (blue squares). The $V_{\text{th}}(n_{\text s})$ dependence extrapolates to zero at the same electron density as $\Delta/k_{\text{B}}$. The same analysis, yielding similar results, has been previously made in a 2D electron system in Si MOSFETs \cite{shashkin2001metal}, thus adding confidence that the MIT in 2D is a genuine quantum phase transition.

We now compare the results for $n_{\text{c}}$ with the behavior of the effective electron mass $m_{\text{F}}$ at the Fermi level (the latter was determined by the analysis of the Shubnikov-de~Haas oscillations; the detailed procedure of measuring $m_{\text{F}}$ is described in Ref.~\cite{melnikov2017indication}). In Fig.~\ref{fig2}(b), the product $n_{\text{s}}g_0m_0/g_{\text{F}}m_{\text{F}}$ is plotted as a function of $n_{\text s}$. Here $g_0$=2 is the Land\'e $g$-factor in the bulk silicon, $m_0=0.19\, m_{\text{e}}$ is the band mass, $m_{\text{e}}$ is the free electron mass, and $g_{\text{F}}\approx g_0$ is the $g$-factor at the Fermi level; for more on this, see section 5. The inverse effective mass extrapolates linearly to zero at a density $n_{\text{m}}=1.1\pm0.1\times10^{10}$~cm$^{-2}$ that is noticeably higher than $n_{\text{c}}$. This is in contrast to the situation in Si MOSFETs, where a similar dependence of the inverse effective mass on the electron density has been observed, but $n_{\text{m}}$ has always been slightly below $n_{\text{c}}$ \cite{shashkin2002sharp,mokashi2012critical}. A natural conclusion is that as the residual disorder in a 2D electron system is decreased, the critical electron density $n_{\text{c}}$, affected by the residual disorder, becomes lower than the density $n_{\text{m}}$, at which the effective mass at the Fermi level tends to diverge. This indicates that these two densities are not directly related.

In closing this section, we would like to stress that the behavior of the electron system in SiGe/Si/SiGe quantum wells is qualitatively different from that in Si MOSFETs. Since the critical electron density in Si MOSFETs always lies slightly above $n_{\text{m}}$, the MIT in this system occurs in a strongly-interacting, but conventional Fermi liquid state.  In contrast, in ultraclean SiGe/Si/SiGe quantum wells, the opposite relation $n_{\text{c}}<n_{\text{m}}$ holds, and the MIT occurs in an unconventional Fermi liquid state at electron density below the topological phase transition expected at $n_{\text s}=n_{\text{m}}$, where the Fermi surface breaks into several separate surfaces \cite{zverev2012microscopic}. This should strengthen the metallic temperature dependence of the resistance \cite{zala2001interaction}, which is consistent with the observation of the low-temperature drop in the resistance by a factor of 12, the highest value reported so far in any 2D system.

\section{Metallic state in a strongly interacting spinless two-valley electron system}\vspace{-5mm}

The existence of the $B=0$ metallic state and the MIT in 2D is intimately related to the presence of spin and valley degrees of freedom in the electron spectrum \cite{punnoose2001dilute,punnoose2005metal,lee1985disordered,fleury2008many,burmistrov2008electronic}. Once the electron spins in a single-valley 2D system become fully polarized by an external magnetic field, the system was predicted to become insulating \cite{lee1985disordered}. On the other hand, the electron spectrum in silicon-based 2D systems contains two almost degenerate alleys, which should promote metallicity \cite{punnoose2001dilute,punnoose2005metal,fleury2008many}. Therefore, the metallic state may, in principle, survive in these systems in the presence of spin-polarizing magnetic fields.

\begin{figure}
\scalebox{.55}{\includegraphics{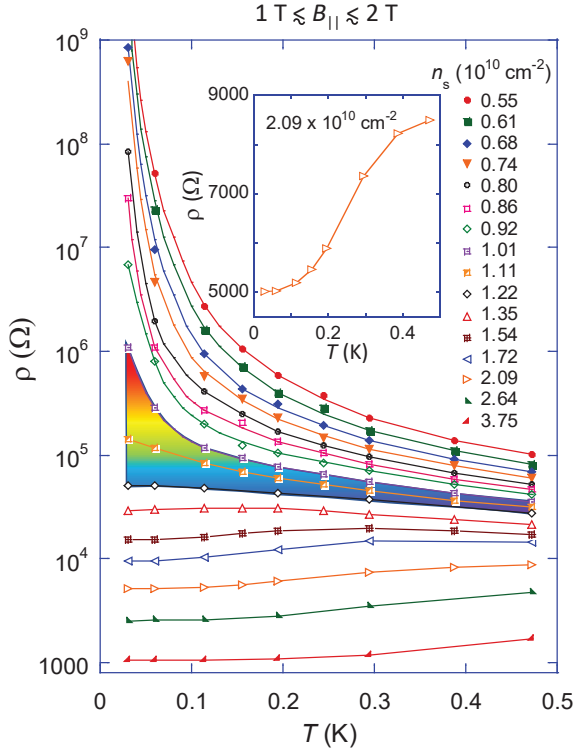}}
\caption{Resistivity of an electron system in a SiGe/Si/SiGe quantum well placed in the spin-polarizing magnetic field $B^*$ as a function of temperature for different electron densities.  The critical region near the MIT is color-gradated.  The magnetic fields used are spanned in the range between approximately 1 and 2~T.  The inset shows a close-up view of $\rho(T)$ for $n_{\text s}=2.09\times10^{10}$~cm$^{-2}$.  From Ref.~\cite{melnikov2020metallic}.\label{fig3}}
\end{figure}

In Fig.~\ref{fig3}, we plot $\rho(T)$ dependences, measured in an ultraclean SiGe/Si/SiGe quantum well at different electron densities in magnetic fields $B_\parallel$ parallel to the 2D plane and strong enough to polarize the electron spins.  The magnetic field of the complete spin polarization, $B^*$, is density-dependent and has been determined by the saturation of the $\rho(B_\parallel)$ dependence, which corresponds to the lifting of the spin degeneracy \cite{okamoto1999spin,vitkalov2000small}.  The values of magnetic fields used in the experiments of Ref.~\cite{melnikov2020metallic} fell within the range between approximately 1 and 2~T. As shown in Fig.~\ref{fig3}, at the lowest temperatures, the resistivity has a strong metallic temperature dependence ($d\rho/dT>0$) at electron densities above a specific critical value, $n_{\text c}(B^*)$, and an insulating behavior ($d\rho/dT<0$) at lower densities.  Assuming that the extrapolation of $\rho(T)$ to zero temperature is valid and taking into account that the dependence separating the metallic and insulating regimes should be tilted \cite{punnoose2005metal}, one can identify the critical density for the MIT at $n_{\text c}(B^*)=(1.11\pm0.05)\times10^{10}$~cm$^{-2}$. At electron densities just above the critical value, the $\rho(T)$ dependences on the metallic side of the transition are non-monotonic: at temperatures exceeding a density-dependent value $T_{\text {max}}$, the derivative $d\rho/dT$ is negative, but it changes sign at $T<T_{\text {max}}$. The measurements in Ref.~\cite{melnikov2020metallic} were restricted to 0.5~K, the highest temperature at which the saturation of the $\rho(B_\parallel)$ dependences could still be achieved.  This restriction is likely to reflect the degeneracy condition for the dilute electron system with low Fermi energy.

The strongest resistivity drop with decreasing temperature below 0.5~K on the metallic side of the transition reaches almost a factor of 2 (see the inset to Fig.~\ref{fig3}), which is weaker compared to the factor of more than 12 drop in this system at $B=0$ (Fig.~\ref{fig1}).  Nevertheless, the metallic temperature behavior of spinless electrons in SiGe/Si/SiGe quantum wells remains substantial and comparable to that observed in $p$-type GaAs/AlGaAs heterostructures in zero magnetic field \cite{hanein1998the,gao2006spin}.

\begin{figure}
\scalebox{.6}{\includegraphics{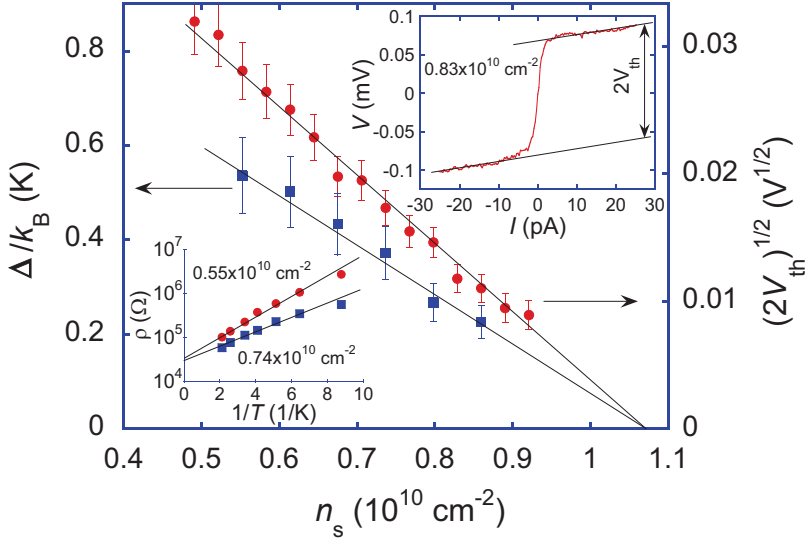}}
\caption{Main panel: The activation energy, $\Delta$, and the square root of the threshold voltage, $V_{\text{th}}^{1/2}$, \textit{vs}.\ electron density.  Solid lines correspond to the best linear fits.  Upper inset: a typical $I$-$V$ dependence on the insulating side of the MIT at $T=30$~mK.  Lower inset: Arrhenius plots of the temperature dependence of the resistivity for two electron densities on the insulating side. From Ref.~\cite{melnikov2020metallic}.\label{fig4}}
\end{figure}

Similarly to the way it was done in the previous section, one can deduce the critical density for the MIT from two additional criteria that do not require the extrapolation of the data to $T=0$: vanishing of the activation energy and nonlinearity of the $I$-$V$ characteristics on the insulating side of the transition. At $n_{\text s}<n_{\text c}(B^*)$, in the vicinity of the critical point, the temperature dependences of the resistivity have an activation character (see the lower inset to Fig.~\ref{fig4}); the density dependence of the activation energy $\Delta$ is plotted in the main panel of Fig.~\ref{fig4}. The dependence is linear and extrapolates to zero at the critical density $n_{\text c}(B^*)=(1.07\pm0.03)\times10^{10}$~cm$^{-2}$. Within the experimental uncertainty, this value coincides with $n_{\text c}(B^*)$ determined from the temperature derivative criterion.  A typical $I$-$V$ characteristic measured on the insulating side of the MIT ($n_{\text s}<n_{\text c}(B^*)$) is shown in the upper inset to Fig.~\ref{fig4}.  The $V(I)$ dependence obeys Ohm's law in a very narrow interval of currents $\left|I\right|\lesssim1$~pA and almost saturates at higher currents.  The square root of $V_{\text{th}}$ is a linear function of $n_{\text s}$ and  extrapolates to zero at the same electron density as the $\Delta(n_{\text s})$ dependence.

As mentioned in the previous section, in zero magnetic field, the independent criteria yield the same critical electron density for the MIT in SiGe/Si/SiGe quantum wells.  This is also the case in Si MOSFETs at $B=0$ \cite{shashkin2001metal}. However, fully spin-polarized (or ``spinless'') electrons behave differently in these two systems.  This difference can be attributed to different intervalley scattering rates.  In Si MOSFETs, where the level of the short-range disorder is some two orders of magnitude higher than that in the ultraclean SiGe/Si/SiGe quantum wells, strong intervalley scattering mixes two valleys at low temperatures effectively producing a single valley \cite{punnoose2010renormalization1,punnoose2010renormalization2,punnoose2010test}, and the derivative criterion fails to yield the critical density for spinless electrons. However, the second criterion mentioned above holds, and this leaves uncertain the existence of the MIT in this system \cite{shashkin2001metal}.  In contrast, in the ultraclean SiGe/Si/SiGe quantum wells, the metallic temperature dependence of the resistivity remains strong even when the electron spins are completely polarized, and both above-mentioned criteria yield the same critical density confirming the existence of the MIT in this 2D system of spinless electrons that retain another, valley degree of freedom. The strength of the metallic temperature dependence of the resistivity is comparable to that in spin-unpolarized single-valley 2D systems in the least disordered $p$-type GaAs/AlGaAs heterostructures, which indicates that the role of distinct valleys in the electron spectrum is equivalent to the role of spins in regard to the existence of the metallic state and the MIT in 2D.

The critical electron density for the MIT in the spinless electron system in SiGe/Si/SiGe quantum wells exceeds that measured in zero magnetic field by a factor of approximately 1.2.  This increase is consistent with the theoretical calculations \cite{dolgopolov2017spin}.  According to this theory, the parallel-field-induced increase in the critical electron density for the Anderson transition in a strongly interacting 2D electron system is due to the exchange and correlation effects; the ratio between the critical electron densities for fully spin-polarized and unpolarized electron systems is independent of the density of impurities and equal to $\approx1.33$.  A similar, although somewhat stronger, suppression of the metallic regime was previously reported in Si MOSFETs where the localization of fully spin-polarized electrons occurs at the electron density by a factor of about 1.4 higher compared to that in zero magnetic field \cite{shashkin2001metal,dolgopolov1992properties,eng2002effects,jaroszynski2004magnetic}.

\section{Manifestation of strong correlations in transport in ultra-clean SiGe/Si/SiGe quantum wells}

Early theories of the metallic state in strongly interacting 2D systems \cite{finkelstein1983influence,finkelstein1984weak,castellani1984interaction} were focused on the interplay between disorder and interactions using renor\-ma\-li\-zation-group scaling theory.  Later, the theory was extended to account for the existence of multiple valleys in the electron spectrum \cite{punnoose2001dilute,punnoose2005metal}.  At temperatures well below the Fermi temperature, the resistivity was predicted to grow with decreasing temperature, reach a maximum at $T=T_{\text{max}}$, and then decrease as $T\rightarrow0$.  The maximum in $\rho(T)$ dependence corresponds to the temperature at which the interaction effects become strong enough to stabilize the metallic state and overcome the quantum localization.  This theoretical prediction, which is applicable only within the so-called diffusive regime (roughly, $k_{\text B}T\tau/\hbar<1$, where $\tau$ is the mean-free time), was found to be consistent with the experimental $\rho(T)$ data in silicon MOSFETs \cite{punnoose2001dilute,punnoose2010test,anissimova2007flow}, but only in a narrow range of electron densities near $n_{\text c}$.  However, strong temperature dependence of the resistivity has been experimentally observed in a wide range of electron densities: up to five times the critical density, including the so-called ballistic regime (roughly, $k_{\text B}T\tau/\hbar>1$), where the renormalization-group scaling theory is not relevant.\footnote{We emphasize that the ballistic regime introduced in Ref.~\cite{zala2001interaction} is not related to the well-known ballistic transport, or Knudsen regime, where the mean free path is larger than the sample dimensions.}

A similar physical mechanism --- the elastic but temperature-dependent scattering of electrons by the Friedel oscillations --- works in principle in both diffusive and ballistic regimes \cite{zala2001interaction}.  The interaction corrections to the conductivity in the corresponding limits have different forms.  In the diffusive regime, they are logarithmic-in-$T$, as follows from the renormalization-group scaling theory for diffusion modes
\cite{finkelstein1983influence,finkelstein1984weak,castellani1984interaction,punnoose2001dilute,punnoose2005metal,lee1985disordered,castellani1998metallic}.  In the ballistic regime, the corrections are linear-in-$T$, according to earlier theories of temperature-dependent screening of the impurity potential \cite{stern1980calculated,gold1986temperature,dassarma1986theory,dassarma1999charged}, where the leading term has the form $\sigma(T)-\sigma(0)\propto T/T_{\text F}$ (note that the Fermi temperature $T_{\text F}$ is in general determined by the effective electron mass $m$ renormalized by interactions).\footnote{The behaviors of the effective electron mass at the Fermi level and the energy-averaged effective electron mass are qualitatively different at low electron densities in the strongly correlated 2D system in SiGe/Si/SiGe quantum wells (see section 5). For the sake of simplicity, in this section we will disregard this difference.} The theory of interaction corrections \cite{zala2001interaction} and the screening theory \cite{gold1986temperature} in its general form, which takes account of the renormalization of the mass, allowed one to extract the effective mass from the slope of the linear-in-$T$ correction to the conductivity in the ballistic regime \cite{shashkin2002sharp,shashkin2004comment}.  In Ref.~\cite{shashkin2002sharp}, it was shown that the so-obtained effective mass sharply increases with decreasing electron density and that the $m(n_{\text s})$ dependence practically coincides with that obtained by alternative measurement methods \cite{shashkin2003spin,anissimova2006magnetization}.  However, the small corrections calculated in the ballistic regime cannot convincingly explain the experimentally observed order-of-magnitude changes in the resistivity with temperature. In principle, in line with the screening theories \cite{gold1986temperature,dassarma1999charged}, one can expect the resistivity to be a function of $T/T_{\text F}$ with a maximum at $T_{\text {max}}\sim T_{\text F}$, above which the electrons are not degenerate.  As of now, there are no accepted theoretical calculations allowing for a quantitative comparison with experiments.

An alternative interpretation of the temperature dependence of the resistivity is based on the so-called Wigner-Mott scenario, which focuses on the role of strong electron-electron interactions. The simplest theoretical approach to non-perturbatively tackle the interactions as the main driving force for the MIT is based on dynamical mean-field theory (DMFT) methods \cite{camjayi2008coulomb,radonjic2012wigner,dobrosavljevic2017wigner} using the Hubbard model at half-filling.  On the metallic side near the MIT, the resistivity is predicted to initially increase as the temperature is reduced, reach a maximum, $\rho_{\text {max}}$, at temperature $T_{\text {max}}\sim T_{\text F}$, and then decrease as $T\rightarrow0$.  It has also been shown that the resistivity change $\rho(T)-\rho(0)$, normalized by its maximum value, is a universal function of $T/T_{\text {max}}$.

Yet another approach to treat the strongly-interacting 2D electron systems was proposed in Refs.~\cite{spivak2003phase,spivak2004phases,spivak2006transport}. It is based on the Pomeranchuk effect expected within a phase coexistence region between the Wigner crystal and a Fermi liquid. The predicted $\rho(T)$ dependences are also non-monotonic: the resistivity increases with decreasing temperature at $T\gtrsim T_{\text F}$ and decreases at lower temperatures.  To the best of our knowledge, currently there is no theory allowing for a quantitative comparison with experiments.

\begin{figure*}
\scalebox{.43}{\includegraphics{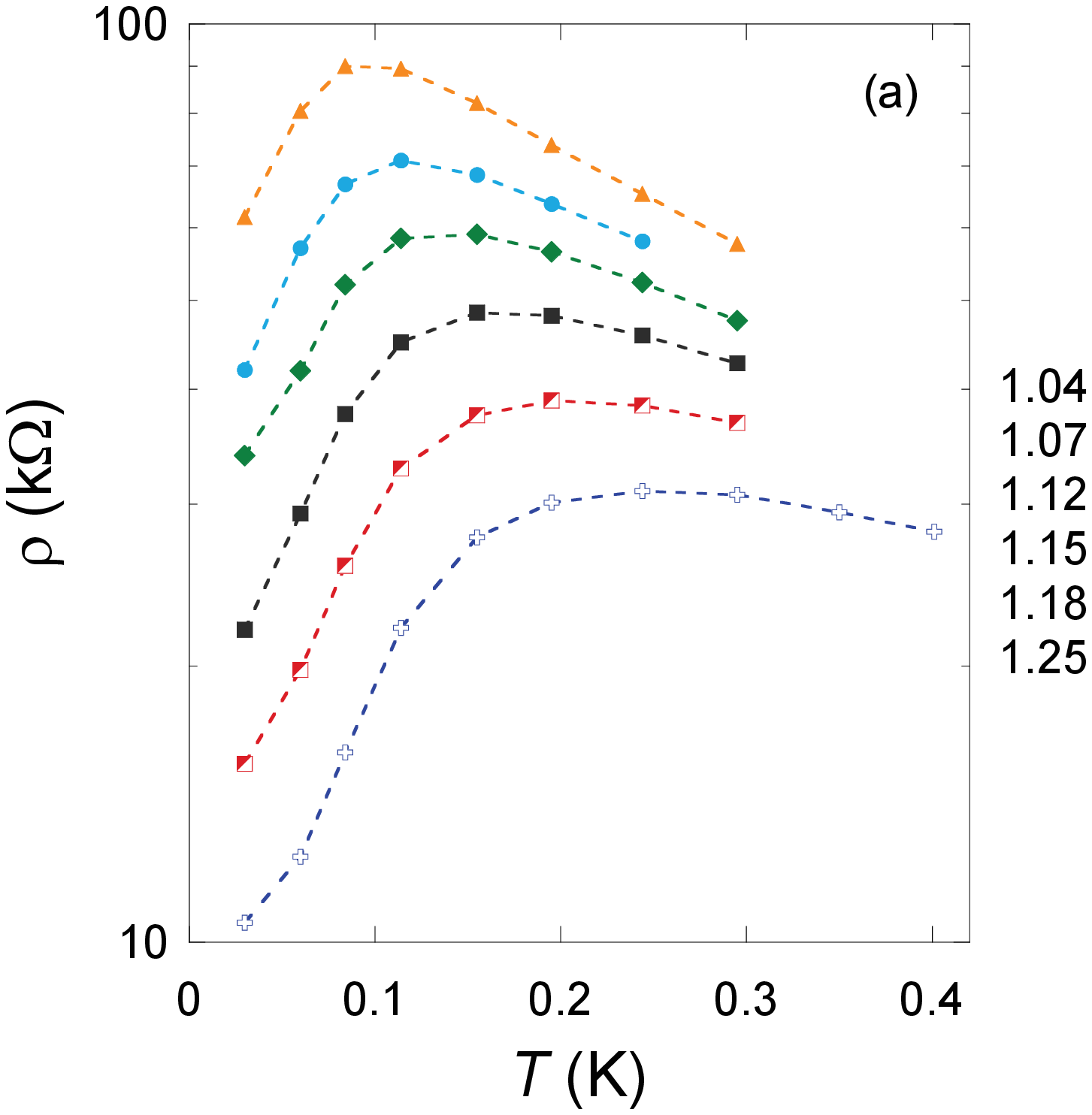}}
\scalebox{.4057}{\includegraphics{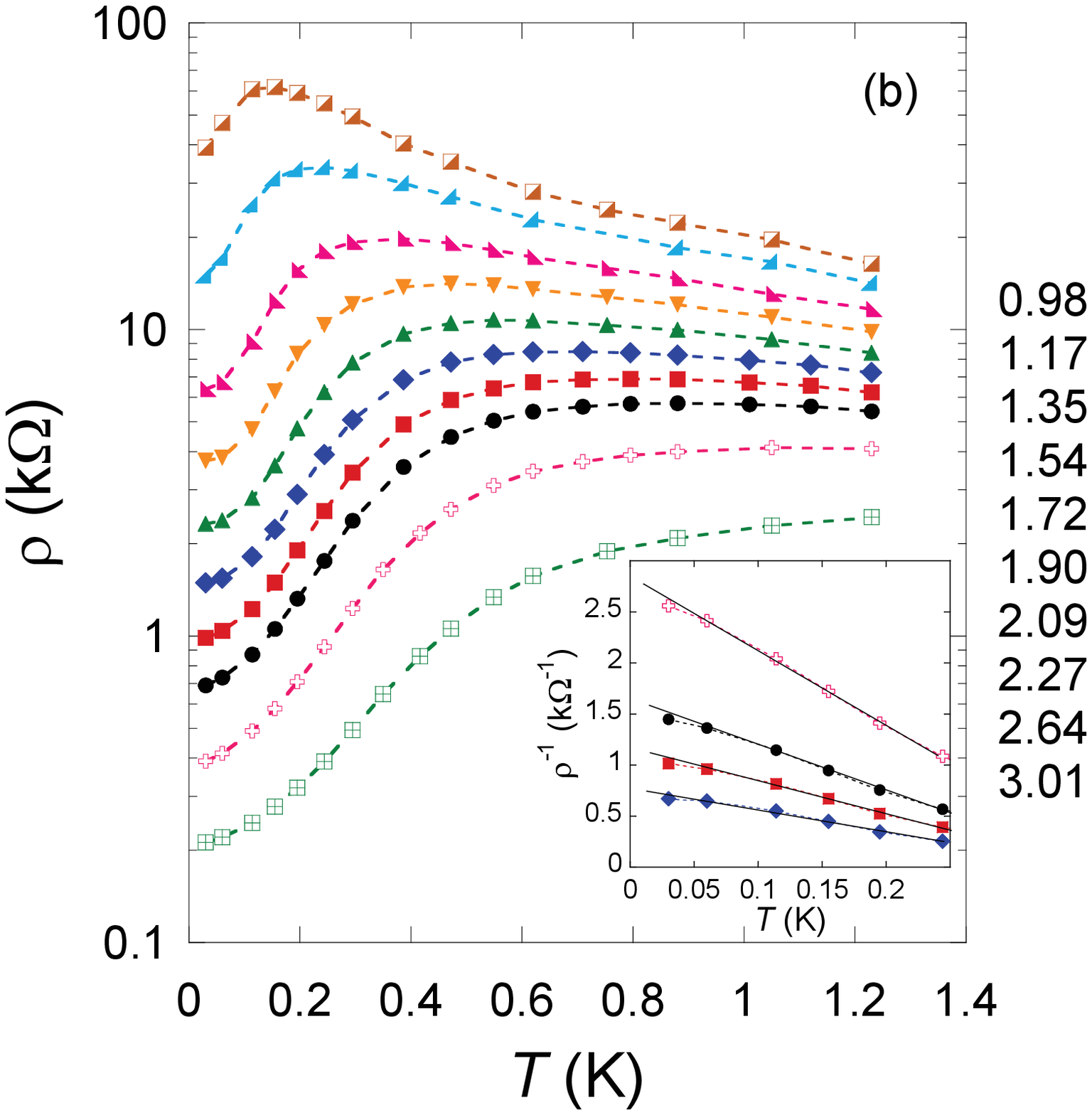}}
\caption{Non-monotonic temperature dependences of the resistivity of the 2D electron system in SiGe/Si/SiGe quantum wells on the metallic side near
the metal-insulator transition for samples A~(a) and B~(b).  The electron densities are indicated in units of $10^{10}$~cm$^{-2}$.  The inset in (b) shows $\rho^{-1}(T)$ dependences for four electron densities in sample B (the symbols are the same as in the main figure).  The solid lines are linear fits to the data.  From Ref.~\cite{shashkin2020manifestation}.\label{fig5}}
\end{figure*}

\begin{figure}
\scalebox{.49}{\includegraphics{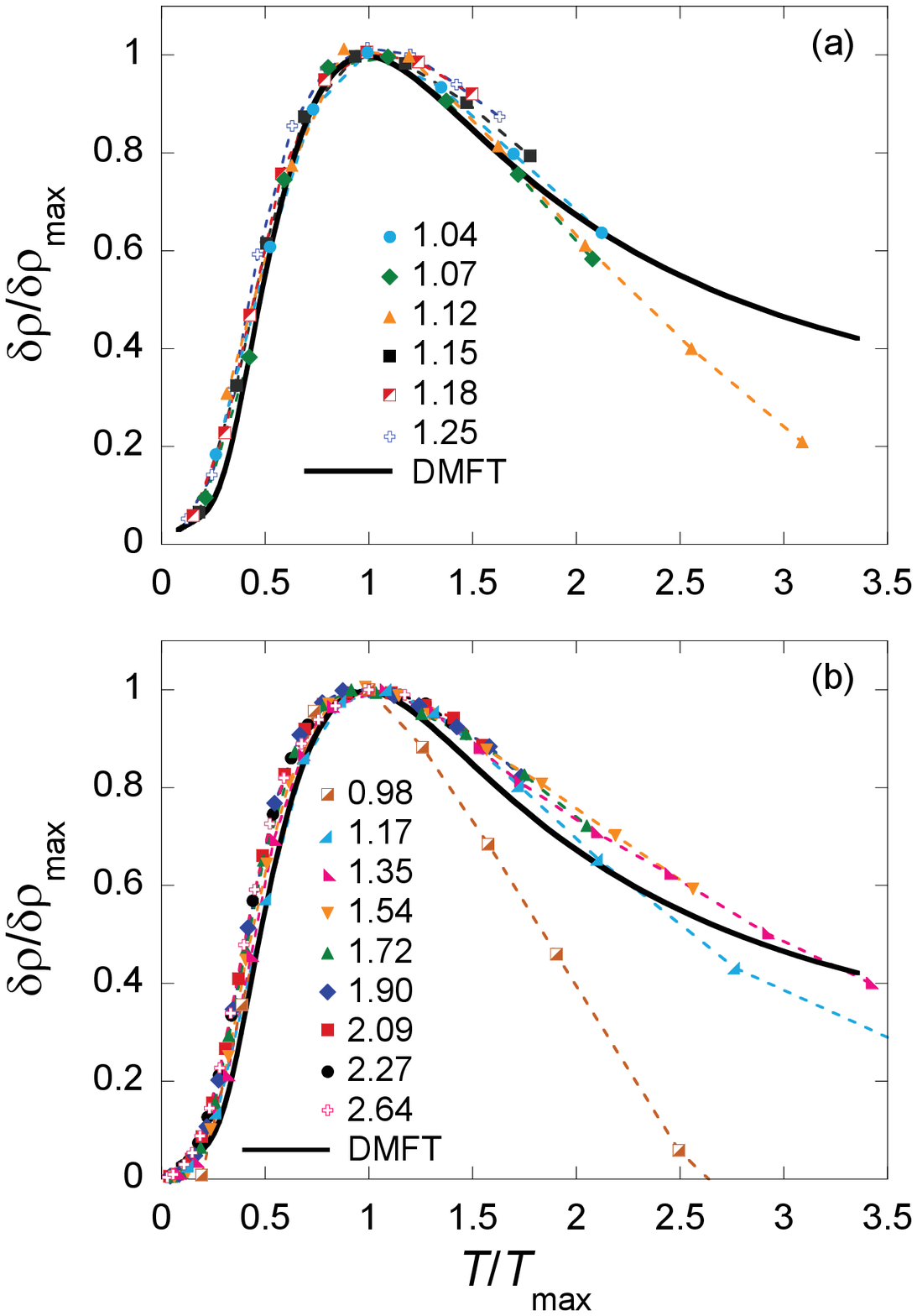}}
\scalebox{.42}{\includegraphics{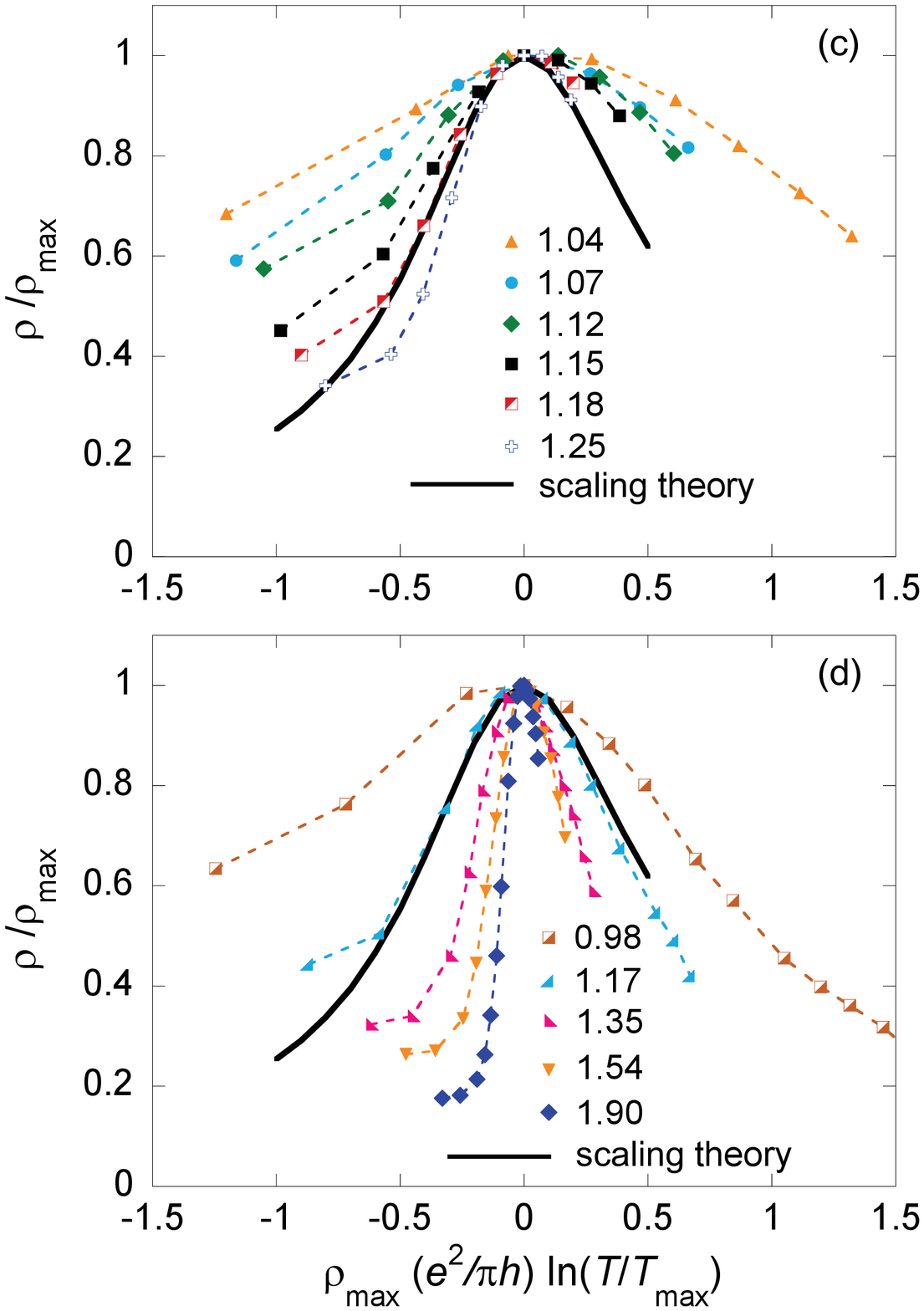}}
\caption{Left-hand side panel: the ratio $(\rho(T)-\rho(0))/(\rho_{\text {max}}-\rho(0))$ as a function of $T/T_{\text {max}}$ for samples A~(a) and B~(b).  Solid lines show the results of DMFT in the weak-disorder limit \cite{camjayi2008coulomb,radonjic2012wigner,dobrosavljevic2017wigner}. Right-hand side panel: the ratio $\rho/\rho_{\text {max}}$ as a function of the product $\rho_{\text {max}} \ln(T/T_{\text {max}})$ for samples A~(c) and B~(d).  Solid lines are the result of the scaling theory \cite{punnoose2001dilute,punnoose2005metal}.  In both panels, the electron densities are indicated in units of $10^{10}$~cm$^{-2}$.  From Ref.~\cite{shashkin2020manifestation}.\label{fig6}}
\end{figure}

In Fig.~\ref{fig5}, temperature dependences of the resistivity in the metallic regime are shown for two SiGe/Si/SiGe samples in the range of electron densities where the $\rho(T)$ curves are non-monotonic: at temperatures below a density-dependent temperature $T_{\text {max}}$, the resistivity exhibits metallic temperature behavior ($\mathrm{d}\rho/\mathrm{d}T>0$), while above $T_{\text {max}}$, the behavior is insulating ($\mathrm{d}\rho/\mathrm{d}T<0$). Note that the resistivity drop at $T<T_{\text {max}}$ in these samples is strong and may exceed an order of magnitude (more than a factor of 12 for the lowest curve in Fig.~\ref{fig5}(b)), which is twice as large compared to that in the best 2D electron systems studied so far. In the inset to Fig.~\ref{fig5}(b), the data recalculated into the conductivity as a function of temperature are plotted. We also show linear fits to the data. The observed linear temperature dependence of conductivity is consistent with the ballistic regime not too close to the critical density. The temperature dependence of the conductivity allows one to conclude that the transient region between ballistic and diffusive regimes corresponds to electron densities around $\approx 1.1\times10^{10}$~cm$^{-2}$.

According to DMFT, the resistivity data should scale when plotted in a form $\delta\rho/\delta\rho_{\text{max}}$ \textit{vs}.\ $T/T_{\text{max}}$ (here $\delta\rho=\rho(T)-\rho(0)$ and $\delta\rho_{\text{max}}=\rho_{\text {max}}-\rho(0)$). The results of the scaling analysis of the data shown in Fig.~\ref{fig5}, performed in Ref.~\cite{shashkin2020manifestation}, are presented in Fig.~\ref{fig6}.  The data scale perfectly in a wide range of $n_{\text s}$ and are described well by the theory (the solid curve) in the weak-disorder limit. We emphasize that at some electron densities, the changes of the resistivity with temperature exceed an order of magnitude.  Deviations from the theoretical curve become pronounced at $T>T_{\text {max}}$ at electron densities within $\sim10$\% of the critical value $n_{\text c}\approx 0.88\times10^{10}$~cm$^{-2}$. The fact that in the low-temperature limit, the same data display linear-in-$T$ corrections to the conductivity (see the inset to Fig.~\ref{fig5}(b)) reveals the consistency of the DMFT and both the theory of interaction corrections \cite{zala2001interaction} and the generalized screening theory \cite{shashkin2004comment}. We argue that the DMFT can be applied to strongly interacting 2D electron systems: the Friedel oscillations near the impurities in real electron systems, even weakened by strong electron correlations \cite{andrade2010quantum}, imply the existence of a short-range spatial charge order that plays the role of an effective lattice. Note that the theory also quantitatively describes weaker non-monotonic $\rho(T)$ dependences in silicon MOSFETs and $p$-GaAs heterostructures \cite{radonjic2012wigner,dobrosavljevic2017wigner}.

We also scale the $\rho(T)$ data in the spirit of the renormalization-group scaling theory \cite{punnoose2001dilute,punnoose2005metal}, according to which, the normalized resistivity $\rho/\rho_{\text {max}}$ is expected to be a universal function of the product $\rho_{\text {max}}\ln(T/T_{\text {max}})$.  The results are plotted in the right-hand panel of Fig.~\ref{fig6}(c,d).  In both samples, only the data obtained at $n_{\text s}=1.18\times10^{10}$~cm$^{-2}$ for sample A (Fig.~\ref{fig6}(c)) and at $n_{\text s}=1.17\times10^{10}$~cm$^{-2}$ for sample B (Fig.~\ref{fig6}(d)) coincide well with the theoretical curve, although some deviations occur at the lowest temperature.  Pronounced deviations from the theory arise at both higher and lower $n_{\text s}$.  At lower electron densities, the scaled experimental curves become wider than the theoretical one, while at higher densities, they become narrower.  A similar shrinkage of the scaled curves with increasing $n_{\text s}$ was reported earlier in Refs.~\cite{punnoose2001dilute,anissimova2007flow,radonjic2012wigner}, where the resistivity data obtained in Si MOSFETs were analyzed.  One should take into account, however, that theory \cite{punnoose2001dilute,punnoose2005metal} has been developed for 2D electron systems that, on the one hand, are in the diffusive regime and, on the other hand, their resistivities are low compared to $\pi h/e^2$ because at higher values of $\rho$, higher-order corrections become important and cause deviations from the universal scaling curve.  As a result, the applicable range of parameters becomes very narrow. Note that an attempt to scale the resistivity data in Si MOSFETs in the spirit of the renormalization-group scaling theory was made in Refs.~\cite{knyazev2007critical,knyazev2008metal} at high temperatures above 1~K for the moderate change in $\rho(T)$. Thus, the $\rho(T)$ data are best described by DMFT.

A question naturally arises of how DMFT and the renormalization-group scaling theory are connected.  Although both theories predict non-monotonic temperature dependences of the resistivity, within the scaling theory \cite{punnoose2001dilute,punnoose2005metal}, the maximum in the $\rho(T)$ dependences occurs at the temperature well below $T_{\text F}$, at which the temperature-dependent interactions become strong enough to overcome the effect of the quantum localization.  This theory is relevant only in the diffusive regime.  In contrast, within the DMFT, the maximum in $\rho(T)$ dependences corresponds to the quasiparticle coherence temperature $T^\ast\sim T_{\text F}$, below which the elastic electron-electron scattering corresponds to coherent transport, while at higher temperatures the inelastic electron-electron scattering becomes strong and gives rise to a fully incoherent transport.  Even though the theoretical estimates of the positions of the maxima may be crude, the origins of the maxima are clearly different within these two theories in view of the role of the disorder. On the other hand, the functional forms of $\rho(T)$ dependences, including the maximum at $T_{\text {max}}\sim T_{\text F}$, expected from both the screening theory in its general form and DMFT, are similar. In particular, the linear temperature dependence of the conductivity at $T\ll T_{\text{F}}$ following from the generalized screening theory \cite{shashkin2004comment} and from the theory of the corrections to the conductivity due to the scattering on Friedel oscillations in the ballistic regime \cite{zala2001interaction} is consistent with the prediction of the DMFT.  This similarity adds confidence in both theories and gives a hint that the underlying microscopic mechanism may be the same, \textit{i.e.}, electron-impurity or impurity-mediated electron-electron scattering in the strongly interacting case.

Finally, we mention that similar non-monotonic $\rho(T)$ dependences are observed \cite{limelette2003mott,kurosaki2005mott} in quasi-two-dimensional organic charge-transfer salts (so-called Mott organics), as well as in 2D transition metal dichalcogenides \cite{moon2020quantum,moon2021metal,li2021continuous}.  Interestingly, DMFT is capable of quantitatively describing $\rho(T)$ dependences in these systems \cite{dobrosavljevic2017wigner,moon2020quantum}, which demonstrates that this theory is applicable to various strongly correlated systems.

\section{Indication of band flattening at the Fermi level}

Flat band materials have recently attracted much attention \cite{heikkila2011flat,bennemann2013novel,peotta2015superfluidity,volovik2015from,khodel2020metamorphoses}.  The interest is caused, in particular, by the fact that, due to the anomalous density of states, the flattening of the band may be important for the construction of room temperature superconductivity.  The formation of a flat band at the Fermi level was theoretically predicted \cite{camjayi2008coulomb,amusia2015theory,yudin2014fermi} in heavy fermions, high-temperature superconducting materials, $^3$He, and two-dimensional electron systems.  As the strength of fermion-fermion interactions increases, the single-particle spectrum becomes progressively flatter in the vicinity of the Fermi energy, eventually forming a plateau. The flattening of the spectrum is related to the increase of the effective fermion mass $m_{\text F}$ at the Fermi level and the corresponding peak in the density of states.

Experimental data obtained in strongly interacting 2D electron systems can be divided into two groups: the data describing the electron system as a whole (for example, the magnetic field of the complete spin polarization) and the data related solely to the electrons at the Fermi level (like the amplitude of the Shubnikov-de~Haas oscillations that yields the effective mass $m_{\text F}$ and Land\'e $g$-factor $g_{\text F}$ at the Fermi level). The results for the energy-averaged values $m$ and $g$ in the first group often turn out to be identical to the results for $m_{\text F}$ and $g_{\text F}$. For example, simultaneous increase of the energy-averaged effective mass and that at the Fermi level was reported in strongly correlated 2D systems in Si~MOSFETs \cite{kravchenko2004metal,shashkin2005metal,pudalov2006metal,shashkin2002sharp,mokashi2012critical,dolgopolov2015two,kuntsevich2015strongly}.  The strongly enhanced effective mass in Si~MOSFETs was previously interpreted in favor of the formation of the Wigner crystal or an intermediate phase (\textit{e.g}., a ferromagnetic liquid). The origin and existence of possible intermediate phases preceding the formation of the Wigner crystal can depend on the level of disorder in the electron system. Since in SiGe/Si/SiGe quantum wells, the electron mobility is some two orders of magnitude higher than that in Si MOSFETs, the origin of the low-density phases in these electron systems can be different. Note that the experimental results obtained in the least-disordered Si MOSFETs exclude localization-driven MIT. The effects of the disorder in higher mobility SiGe/Si/SiGe quantum wells should be yet weaker.

In this section, we compare the energy-averaged product $g_{\text F}m$ and the product $g_{\text F}m_{\text F}$ at the Fermi level in ultra-clean SiGe/Si/SiGe quantum wells.  The magnetic field of the complete spin polarization, $B^*(n_{\text s})$, which corresponds to a distinct ``knee'' on the magnetoresistance curves \cite{okamoto1999spin,vitkalov2000small}, is in good agreement with the theoretical dependence calculated using the quantum Monte Carlo method for the clean limit $k_{\text F}l\gg 1$ \cite{fleury2010energy} (here $k_{\text F}$ is the Fermi wavevector and $l$ is the mean free path). The product $g_{\text F}m$ can be obtained in the clean limit from the analysis of the measured $B^*(n_{\text s})$ dependence.\footnote{Note that in the case of strong disorder potential, the experimental dependence $B^*(n_{\text s})$ is shifted to higher electron densities due to the presence of localized electron moments \cite{dolgopolov2002comment,gold2002on,prus2003thermodynamic,teneh2012spin,pudalov2018probing}.}  At $B_\parallel=B^*$, the Zeeman splitting is equal to the Fermi energy of the spin-polarized electron system
\begin{equation}
g_{\text F}\mu_{\text B}B^*=\frac{2\pi\hbar^2n_{\text s}}{mg_{\text v}},\label{gm}
\end{equation}
where $\mu_{\text B}$ is the Bohr magneton.  On the other hand, the Land\'e $g$-factor $g_{\text F}$ and effective mass $m_{\text F}$ at the Fermi level can be determined by the analysis of the Shubnikov-de~Haas oscillations in relatively weak magnetic fields. For the details on how both masses have been extracted, see Ref.~\cite{melnikov2017indication}.

\begin{figure}
\scalebox{.5}{\includegraphics{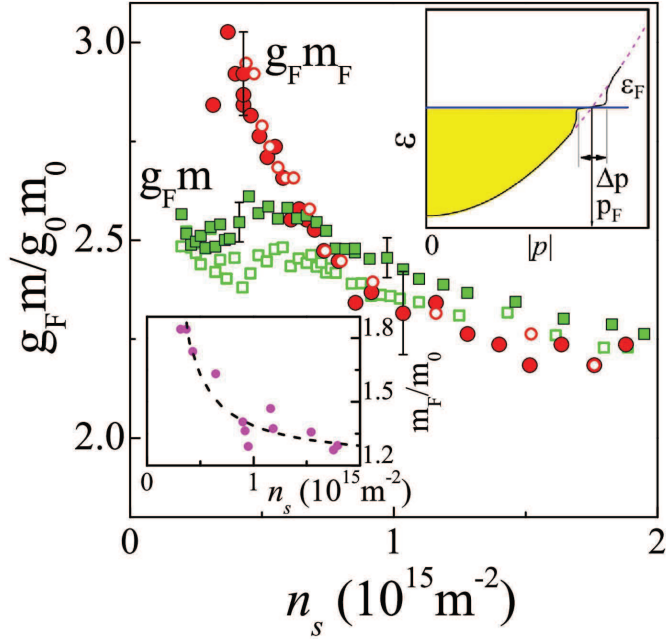}}
\caption{Product of the Land\'e factor and effective mass as a function of electron density determined by measurements of the field of full spin polarization, $B^*$ (squares), and Shubnikov-de~Haas oscillations (circles) at $T\approx 30$~mK. The empty and filled symbols correspond to two samples. The experimental uncertainty corresponds to the data dispersion and is about 2\% for the squares and about 4\% for the circles. ($g_0=2$ and $m_0=0.19\,m_{\text e}$ are the values for noninteracting electrons). The top inset schematically shows the single-particle spectrum of the electron system in a state preceding the band flattening at the Fermi level (solid black line). The dashed violet line corresponds to an ordinary parabolic spectrum. The occupied electron states at $T=0$ are indicated by the shaded area. Bottom inset: the effective mass $m_{\text F}$ versus electron density determined by analysis of the temperature dependence of the amplitude of Shubnikov-de~Haas oscillations.  The dashed line is a guide to the eye. From Ref.~\cite{melnikov2017indication}.\label{fig7}}
\end{figure}

In Fig.~\ref{fig7}, the main result of this section is shown. The average $g_{\text F}m$ and $g_{\text F}m_{\text F}$ at the Fermi level behave similarly at high electron densities, where electron-electron interactions are relatively weak.  However, at low densities, where the interactions become especially strong, their behavior is qualitatively different: the product $g_{\text F}m_{\text F}$ continues to monotonically increase as the electron density is reduced, while the product $g_{\text F}m$ saturates at low $n_{\text s}$. We emphasize that what matters here is the qualitative difference in the behaviors of the two sets of data, rather than a comparison of their absolute values. Since the exchange effects in the 2D electron systems in silicon are negligible \cite{kravchenko2004metal,shashkin2005metal}, one can only attribute this difference to the different behaviors of the two effective masses. Their qualitatively different behavior indicates the interaction-induced band flattening at the Fermi level in this electron system. To add confidence in our results and conclusions, in the bottom inset to Fig.~\ref{fig7} we show the data for the effective mass $m_{\text F}$ determined by the analysis of the temperature-dependent amplitude of the Shubnikov-de~Haas oscillations, as described in Ref.~\cite{melnikov2014effective}. Similar density dependence of $m_{\text F}$ and $g_{\text F}m_{\text F}$ allows one to exclude any possible influence of the $g$-factor on the behavior of the product of the effective mass and the $g$-factor, in consistency with the previous results in silicon MOSFETs.

We interpret these experimental results within the concept of the fermion condensation \cite{zverev2012microscopic,khodel1990superfluidity,nozieres1992properties} that occurs at the Fermi level in a range of momenta, unlike the condensation of bosons. When the strength of the electron-electron interactions increases, the single-particle spectrum flattens in a region $\Delta p$ near the Fermi momentum $p_{\text F}$ (see top inset to Fig.~\ref{fig7}). At relatively high electron densities $n_{\text s}>0.7\times 10^{15}$~m$^{-2}$, this effect is unimportant because the single-particle spectrum does not change noticeably in the interval $\Delta p$, and the behaviors of the energy-averaged effective mass and that at the Fermi level are practically identical. Decreasing the electron density in the range $n_{\text s}<0.7\times 10^{15}$~m$^{-2}$ gives rise to the flattening of the spectrum so that the effective mass at the Fermi level, $m_{\text F}=p_{\text F}/v_{\text F}$, continues to increase (here $v_{\text F}$ is the Fermi velocity). In contrast, the energy-averaged effective mass does not because it is not particularly sensitive to this flattening.

\section{Transport evidence for a sliding two-dimensional quantum electron solid}

Experimental studies \cite{melnikov2017indication,mokashi2012critical,shashkin2001indication,vitkalov2001scaling} of the transport and thermodynamic properties of strongly correlated 2D electron systems have suggested that at low electron densities, these systems approach a phase transition to a new, unknown state that could be a quantum Wigner crystal or a precursor \cite{wigner1934on,chaplik1972possible,tanatar1989ground,attaccalite2002correlation,kagalovsky2020hartree}. (The term quantum means that the kinetic energy of 2D electrons is determined by the Fermi energy in contrast to the classical Wigner crystal \cite{grimes1979evidence}, in which the kinetic energy of electrons is determined by temperature.) The phase transition point in the least-disordered 2D electron systems was found to be close to the critical electron density for the MIT. Although the low-density insulating state has been extensively studied in different 2D systems \cite{andrei1988observation,goldman1990evidence,williams1991conduction,santos1992observation,pudalov1993zero,yoon1999wigner,chitra2005zero,knighton2018evidence,huang2018transport}, no definitive conclusion has been reached about its origin.  While many authors have interpreted the observed nonlinear current-voltage ($I$-$V$) curves as manifestation of the depinning of the Wigner crystal, alternative explanations of the breakdown of the insulating phase have been proposed based on traditional scenarios such as electron heating and subsequent thermal runaway \cite{jiang1991magnetotransport}, as well as Efros-Shklovskii variable range hopping in strong electric field or percolation \cite{shashkin2005metal,marianer1992effective,dolgopolov1992metal}).

\begin{figure}
\scalebox{.57}{\includegraphics{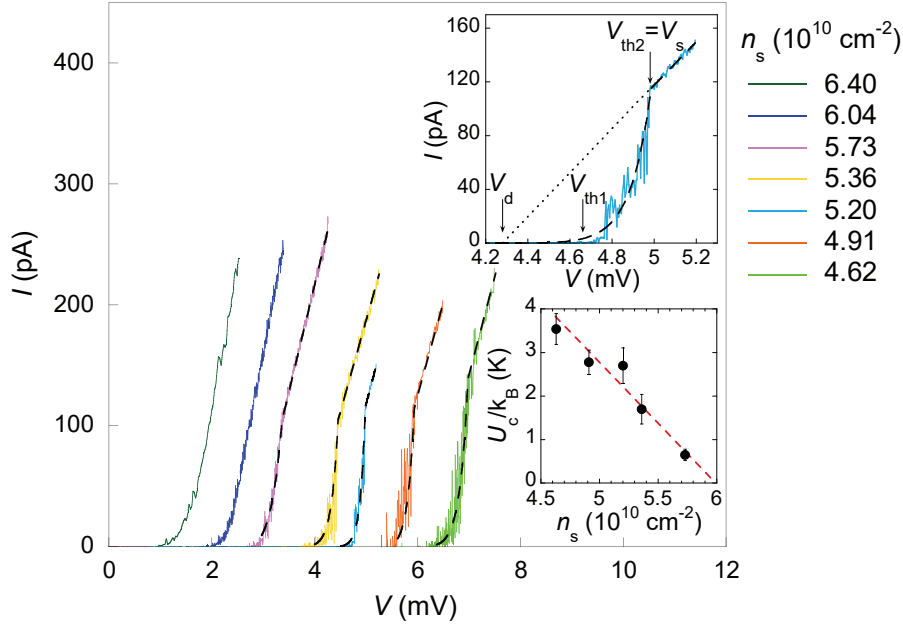}}
\caption{$V-I$ curves are shown for different electron densities in the insulating state at a temperature of 60~mK. The dashed lines are fits to the data using Eq.~(\ref{I}). The top inset shows the $V$-$I$ curve for $n_{\text{s}}=5.20\times 10^{10}$~cm$^{-2}$ on an expanded scale; also shown are the threshold voltages $V_{\text{th1}}$ and $V_{\text{th2}}$, the static threshold $V_{\text{s}}=V_{\text{th2}}$, and the dynamic threshold $V_{\text{d}}$ that is obtained by the extrapolation of the linear region of the $V$-$I$ curve to zero current. Bottom inset: activation energy $U_{\text{c}}$ \textsl{vs}.\ electron density.  Vertical error bars represent standard deviations in the determination of $U_{\text{c}}$ from the fits to the data using Eq.~(\ref{I}).  The dashed line is a linear fit.  From Ref.~\cite{brussarski2018transport}.\label{fig8}}
\end{figure}

In this section, we discuss double-threshold voltage-current characteristics and accompanying noise recently observed in a strongly interacting 2D electron system in silicon MOSFETs at very low electron densities \cite{brussarski2018transport}. In Fig.~\ref{fig8}, a set of low-temperature voltage-current curves, measured at different electron densities in the insulating regime is shown.  The critical electron density for the MIT in this sample is $n_{\text{c}}\approx 8\times 10^{10}$~cm$^{-2}$; the corresponding interaction parameter at this density is $r_{\text{s}}\approx20$. At electron densities below $\approx 6\times 10^{10}$~cm$^{-2}$, two threshold voltages are observed: with increasing bias voltage, the current remains near zero up to the first threshold voltage $V_{\text{th1}}$; then it sharply increases until a second threshold voltage $V_{\text{th2}}$ is reached, above which the slope of the $V$-$I$ curve is significantly reduced, and the dependence becomes linear, although not ohmic (see also the top inset to Fig.~\ref{fig8}). As the electron density is increased, the value of $V_{\text{th1}}$ decreases while the second threshold becomes less pronounced and eventually disappears. No hysteresis was observed in the entire range of electron densities studied.  We emphasize that the observed two-threshold behavior is quite distinct from that reported in the insulating state in amorphous InO films, where the current was found to jump at the threshold voltage by as much as five orders of magnitude and the $V$-$I$ curves exhibited hysteresis consistent with bistability and electron overheating \cite{ovadia2009electron,altshuler2009jumps}.  Furthermore, in the experiments of Ref.~\cite{brussarski2018transport}, the power dissipated near the onset $V_{\text{th1}}$ was less than $10^{-16}$~W, which is unlikely to cause substantial electron overheating, while the power dissipated near the threshold voltage in Ref.~\cite{ovadia2009electron} was more than three orders of magnitude higher. Note also that the double-threshold $V$-$I$ characteristics cannot be explained within the percolation picture according to which, a single threshold is expected \cite{shashkin2005metal}. Thus, the existing traditional mechanisms \cite{shashkin2005metal,jiang1991magnetotransport,marianer1992effective,dolgopolov1992metal} cannot account for the double-threshold behavior reported in Ref.~\cite{brussarski2018transport}.

\begin{figure}
\scalebox{.67}{\includegraphics{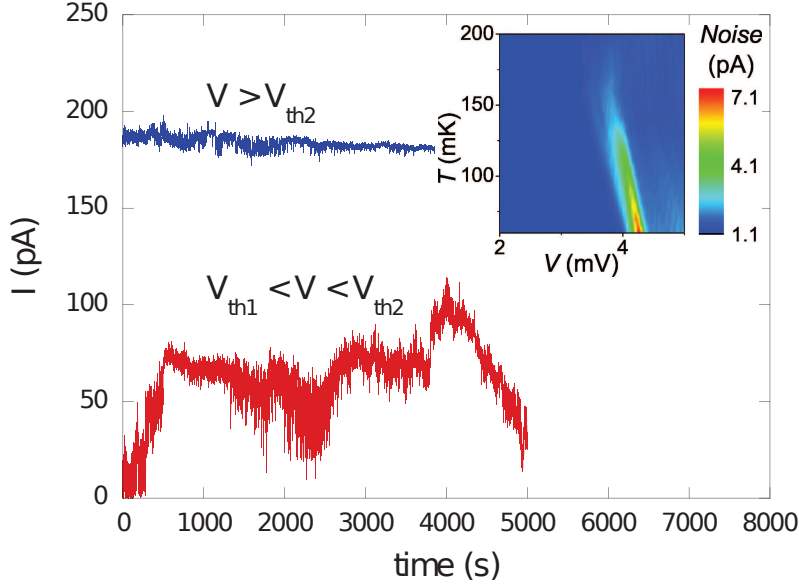}}
\caption{Current is plotted as a function of time for $n_{\text{s}}=5.2\times 10^{10}$~cm$^{-2}$ and $T=60$~mK at voltages $V=4.90$~mV (which lies between $V_{\text{th1}}$ and $V_{\text{th2}}$; lower curve) and $V=5.44$~mV (above $V_{\text{th2}}$). Inset: color map of the broad-band noise at $n_{\text{s}}=5.36\times 10^{10}$~cm$^{-2}$ on a $(V,T)$ plane. From Ref.~\cite{brussarski2018transport}.\label{fig9}}
\end{figure}

It is important that at bias voltages between the two thresholds, the current exhibits strong fluctuations with time that are comparable to its value.  This is shown in Fig.~\ref{fig9}, where the current is plotted as a function of time for density $n_{\text{s}}=5.2\times 10^{10}$~cm$^{-2}$. Above the second threshold, however, these anomalously large fluctuations disappear, and the noise is barely perceptible.

\begin{figure}
\scalebox{.62}{\includegraphics{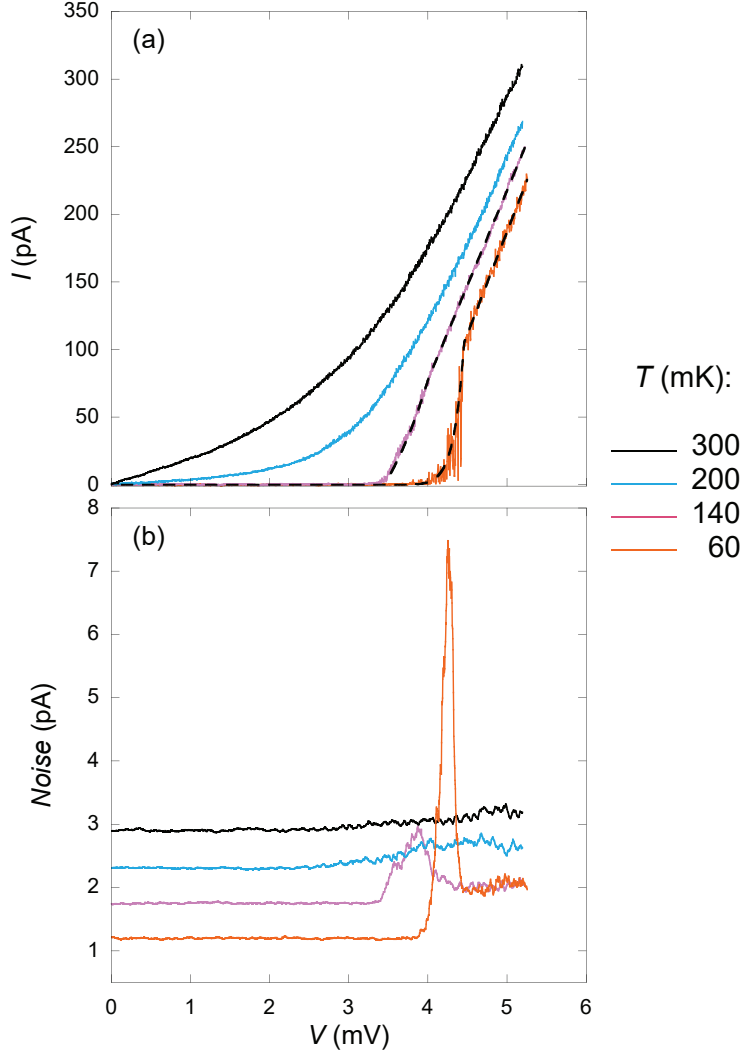}}
\caption{(a) $V-I$ characteristics at $n_{\text{s}}=5.36\times 10^{10}$~cm$^{-2}$ for different temperatures. The dashed lines are fits to the data using Eq.~(\ref{I}). (b)~The broad-band noise as a function of voltage for the same electron density and temperatures.  The three upper curves are shifted vertically for clarity. From Ref.~\cite{brussarski2018transport}.\label{fig10}}
\end{figure}

The measured broad-band noise as a function of voltage is shown in Fig.~\ref{fig10}(b) for different temperatures at electron density $n_{\text{s}}=5.36\times 10^{10}$~cm$^{-2}$. The inset to Fig.~\ref{fig9} is a color map of the broad-band noise on a $(V,T)$ plane. At the lowest temperature, a large increase in the noise is observed between the thresholds $V_{\text{th1}}$ and $V_{\text{th2}}$. This large noise decreases rapidly with increasing temperature in agreement with the two-threshold behavior of the $V$-$I$ curves shown in Fig.~\ref{fig10}(a).

These results have been analyzed in Ref.~\cite{brussarski2018transport} in light of a phenomenological theory based on pinned elastic structures. There is a conspicuous similarity between the double-threshold $V$-$I$ dependences displayed in Fig.~\ref{fig8} and those (with the voltage and current axes interchanged) known for the collective depinning of the vortex lattice in Type-II superconductors (for a comprehensive review, see Ref.~\cite{blatter1994vortices}). The physics of the vortex lattice depinning, in which the existence of two thresholds is well known, was adapted in Ref.~\cite{brussarski2018transport} for the case of an electron solid. In a superconductor, current flows for zero voltage, and the depinning of the vortex lattice occurs when a non-zero voltage appears. Here the situation is reciprocal: a bias voltage is applied but at first the current does not flow in the limit of zero temperature; the depinning of the electron solid is indicated by the appearance of a non-zero current. In the transient region between the dynamic ($V_{\text{d}}$) and static ($V_{\text{s}}$) thresholds, the collective pinning of the solid occurs at the centers with different energies, and the current is thermally activated:
\begin{equation}
I\propto\exp\left[-\frac{U(V)}{k_{\text{B}}T}\right],\nonumber
\label{exp}
\end{equation}
where $U(V)$ is the activation energy. The static threshold $V_{\text{s}}=V_{\text{th2}}$ signals the onset of the regime
of solid motion with friction. This corresponds to the condition
$eEL_{\text{c}}=U_{\text{c}}$, where $E$ is the electric field and $L_{\text{c}}$ is the characteristic distance between the pinning centers with maximal activation energy $U_{\text{c}}$. From the balance of the electric, pinning, and friction forces in the regime of solid motion with friction, one expects a linear $V$-$I$ dependence offset by the threshold $V_{\text{d}}$ corresponding to the pinning force: $I=\sigma_0(V-V_{\text{d}})$, where $\sigma_0$ is a coefficient. Assuming that the activation energy for the electron solid is equal to
\begin{equation}
U(V)=U_{\text{c}}-eEL_{\text{c}}=U_{\text{c}}(1-V/V_{\text{s}}),\nonumber
\label{U}
\end{equation}
one obtains the expression for the current
\begin{equation}
I=\left\{\begin{array}{l}
\sigma_0(V-V_{\text{d}})\exp\left[-\frac{U_{\text{c}}(1-V/V_{\text{s}})}{k_{\text{B}}T}\right] {\text{ for }} V_{\text{d}}<V\leq V_{\text{s}}\\
\sigma_0(V-V_{\text{d}}) {\text{ for }} V>V_{\text{s}}.
\end{array}\right.\label{I}
\end{equation}

In Figs.~\ref{fig8} and \ref{fig10}(a), the fits to the data using Eq.~(\ref{I}) are shown by dashed lines. The experimental two-threshold $V$-$I$ characteristics are described well by Eq.~(\ref{I}). The extracted value of $U_{\text{c}}$ decreases approximately linearly with electron density and tends to zero at $n_{\text{s}}\approx 6\times 10^{10}$~cm$^{-2}$ (see the bottom inset to Fig.~\ref{fig8}). This is in contrast with the vanishing activation energy $\Delta$ of electron-hole pairs at $n_{\text{c}}$ obtained by measurements of the resistance in the limit of zero $I$ and $V$ \cite{shashkin2001metal}; see also section 2. The vanishing $U_{\text{c}}$ is likely to be related to the minimum number of the strong pinning centers for which the collective pinning is still possible. The approximate constancy of the coefficient $\sigma_0\approx1.6\times10^{-7}$~Ohm$^{-1}$ indicates that the motion of the solid with friction is controlled by weak pinning centers \cite{blatter1994vortices}. We argue that the strong noise seen in the regime of the collective pinning of the solid between $V_{\text{d}}$ and $V_{\text{s}}$ should be suppressed in the regime of solid motion with friction at $V>V_{\text{s}}$. Indeed, in the regime of the collective pinning, the solid deforms locally when the depinning occurs at some center, and then this process repeats at another center and so on, leading to the generation of a strong noise. In contrast, in the regime of solid motion with friction, the solid slides as a whole due to the over-barrier motion, and, therefore, the noise is suppressed. Thus, the physics of pinned periodic/elastic objects is relevant for the low-density state in a 2D electron system in silicon MOSFETs. These experimental results are also consistent with numerical simulations of the dynamics of a 2D electron system forming a Wigner solid in the presence of charged impurities \cite{reichhardt2001moving,reichhardt2017depinning}. Although the model proposed in Ref.~\cite{brussarski2018transport} successfully describes the experimental results, further comprehensive theoretical studies are needed.

\section{Conclusions}

We have reviewed recent studies of the MIT and low-density phases in strongly correlated ultraclean silicon-based structures.  Despite much progress has been done in the field, many challenges still remain to be addressed.

\section*{Acknowledgments}
We acknowledge useful discussions with I.~S. Burmistrov, V. Dobrosavljevi\'c, V.~T. Dolgopolov, A.~M. Finkel'stein, D. Heiman, and M.~M. Radonji\'c. A.A.S.\ was supported by RFBR Grant No.\ 19-02-00196 and a Russian Government contract. S.V.K.\ was supported by NSF Grant No.\ 1904051.


\end{document}